\documentclass[12pt]{article}
\hoffset = - 0.5 truecm \voffset=-1.4 truecm
\def\beq{\begin{equation}}   \def\eeq{\end{equation}}
\def\bea{\begin{eqnarray}}  \def\eea{\end{eqnarray}} \def\nn{\nonumber}
\def\noi{\noindent} \def\beeq{\begin{eqnarray}}
\def\eeeq{\end{eqnarray}}
\def\lsim{\raise0.3ex\hbox{$<$\kern-0.75em\raise-1.1ex\hbox{$\sim$}}}
\def\gsim{\raise0.3ex\hbox{$>$\kern-0.75em\raise-1.1ex\hbox{$\sim$}}}
\font \eightrm = cmr10 at 8 pt
\usepackage{epsfig}
\usepackage{graphicx}

\begin{document}

\begin{titlepage}

\begin{flushright}

LPT-ORSAY-04-48

\end{flushright}
\vspace{1.cm}

\begin{center}
\vbox to 1 truecm {}
{\large \bf  Electroproduction Cross Section of Large-E$_{\bot}$ 
Hadrons }\par \vskip 3 truemm
{\large \bf   at NLO and Virtual Photon Structure Function}

\vskip 1 truecm
{\bf M. Fontannaz} \vskip 3 truemm

{\it Laboratoire de Physique Th\'eorique, UMR 8627 CNRS,\\
Universit\'e Paris XI, B\^atiment 210, 91405 Orsay Cedex, France}
\vskip 2 truecm

\begin{abstract}
We calculate Higher Order corrections to the resolved component of the
electroproduction cross section of large-$E_{\bot}$ hadrons. The parton
distributions in the virtual photon are studied in detail and a NLO
parametrization of the latter is proposed. The contribution of the
resolved component to the forward production of large-$E_{\bot}$
hadrons is calculated and its connection with the BFKL cross section is
discussed.
\end{abstract} \end{center}
\end{titlepage}

\pagestyle{plain}
\baselineskip=22 pt
\section{Introduction}
\hspace*{\parindent} The electroproduction cross section of 
large-$E_{\bot}$ hadrons can be split up in two parts. One of
them describes the reaction in which the initial virtual photon takes
part directly in the hard scattering process~; it is called the {\it
direct} part. But the photon can also act as a composite object which is
a source of collinear partons which will take part in the hard
subprocess~; this mechanism is usually refered to as the resolved
process and defines the parton distributions in the {\it virtual}
photon which have the feature of being proportional to $\ell n\
E_{\bot}^2/Q^2$ in the asympototic region where $E_{\bot}^2 \gg Q^2$
(virtually $Q^2$ is the absolute value of the photon).\par

This distinction between direct and resolved component parts is
especially useful in photoproduction reactions in which a quasi-real
photon is present in the initial state (for a review, see ref.
\cite{1r}). In this case the parton distributions in the real photon
are proportional to $\ell n\  E_{\bot}^2/\Lambda_{QCD}^2$ and can be quite
large. The interest in these real distributions dates from the
pioneering work by Witten \cite{2r} who showed that their asymptotic
behavior can be completely calculated in perturbative QCD, a result
which opened the way to interesting tests of the theory. Nevertheless, when
$E_{\bot}^2/\Lambda_{QCD}^2$ decreases, the importance of the non
perturbative contributions grows and we return to a situation similar
to that of the proton structure functions for which non perturbative
inputs are necessary. \par

The situation is clearer when the initial photon is not real, but has a
virtuality $Q^2$ much larger than $\Lambda_{QCD}^2$. In this case the non
perturbative contributions (for instance that of the Vector Meson Dominance
type) are suppressed by powers of $Q^2$ and we are back in the realm of
perturbative QCD. The magnitude of the virtual distributions
is smaller than that of the real distributions. Nonetheless, they are
observable and dedicated experiments have studied the virtual parton
distributions in $e^+e^-$ collisions \cite{3r, 4r} and in the
electroproduction of large $E_{\bot}$ jets \cite{5r,6r,7r} and hadrons
\cite{8r,10rnew}. These studies acquire a quantitative status when data
are compared with theoretical predictions calculated beyond the Leading
Logarithm approximation \cite{10r,12rnew,11r,13new,14new}. It is the 
aim of this paper
to establish such NLO expressions for the resolved component of the
electroproduction of large-$E_{\bot}$ hadrons. We studied the 
corresponding direct
component in ref. \cite{12r}. \par

This work puts the theoretical predictions on a firmer ground since the
full cross section formed by the direct and the resolved component
parts is now calculated at the NLO approximation. In ref. \cite{12r} we
founded predictions for the leptoproduction of forward
large-$E_{\bot}$ hadrons on a NLO calculation of the direct term only.
Then we observed that the resolved component, calculated at the lowest
order, was not negligible. Here we pursue this study of the forward
production now including the HO corrections to the resolved part. This
allows us to refine our predictions and our comparisons with the BFKL-type
cross section which should constitute a non negligible part of the
forward cross section \cite{15new,35r}.

In the next section we gather kinematical definitions and general
expressions concerning the resolved cross section, including a
discussion of the kinematical domain in which such a resolved component
can be defined. Section 3 is devoted to the general structure of the
NLO corrections and the issue of the factorization scheme. In
Section 4 we propose a parametrization of the NLO parton distributions
in the virtual photon, finally, we consider some numerical applications in
Section 5.

\section{The resolved component} \hspace*{\parindent} In this section we
present the kinematical definitions and the general expressions
necessary for the study of the resolved component. This determines the
frame in which the HO calculation described in the next section, will
be performed. The cross section of the reaction $e(\ell ) + p(P) \to e
(\ell ') + h(P_4)+X$,

\beq \label{1e} {d\sigma \over d \varphi dQ^2dy} = {\alpha \over 2
\pi}\ {1 \over 2 \pi}\ {1 \over 2 S} ÷ {1 \over 2} \int {\ell^{\mu\nu}
T_{\mu\nu} \over Q^4} \ dPS \ , \eeq

\noi is written in terms of the leptonic tensor $\ell^{\mu\nu} =
2(\ell^{\mu} \ell '^{\nu} + \ell^{\nu} \ell '^{\mu} - g^{\mu\nu} (\ell \cdot
\ell ' - m_e^2))$ and of the hadronic tensor $T_{\mu\nu}$ which
describes the photon-proton collision. We define the photon variables
$Q^2 = - q^2 = - (\ell - \ell ')^2$ and $y = {q^0 - q^z \over
\ell^0 - l^z} = {P\cdot q \over P \cdot \ell} = 
Q^2/(x_{{\scriptscriptstyle B}_j} S)$
in a frame in which $P^{\mu}$ has no transverse component (we neglect
the proton mass and $P^z$ is positive (HERA convention)). $S$ is given
by $S = (P + \ell)^2$ and $x_{{\scriptscriptstyle B}_j}$ has the 
usual definition $x_{{\scriptscriptstyle B}_j}
= Q^2/2P\cdot q$~; $\varphi$ is the photon azimuthal angle. The
differential phase space of the final hadrons is given by (a sum
over the number of final hadrons is understood in (\ref{1e}))

\beq
\label{2e}
dPS = (2 \pi )^4 \delta^4 \left ( q + P - \sum_{i=1}^n  p_i\right )
\prod_{i=1}^n {d^4p_i
\over (2 \pi )^3} \delta (p_i^2) \theta (p_i^0)\ . \eeq

The hadronic tensor can be calculated as a convolution between the
partonic tensor $t_{\mu\nu}$ which describes the interaction between the
virtual photon and the parton of the proton, and the parton
distribution in the proton $G_a(x,M)$. The fragmentation of the final
parton which produces a large-$E_{\bot}$ hadron is described by the
fragmentation function $D_b^h (z, M_F)$. These distributions depend on
the factorization scales $M$ and $M_F$,

\beq \label{3e} \int T_{\mu\nu} dPS = \sum_{a,b} \int {dx \over x}
G_a(x,M) \int
dz \ D_b^h(z, M_F) t_{\mu\nu}^{ab} \cdot dps \eeq

\noi where $dps$ is the phase space element of the partons produced in
the hard photon-parton collision. From expressions (\ref{2e}) and
(\ref{3e}), we obtain

\bea \label{4e} &&{d\sigma \over d\varphi dQ^2 dydE_{\bot 4}
d\eta_4} = {E_{\bot 4} \over 2 \pi} {\alpha \over 2 \pi} \sum_{a,b}
\int dx G_a(x,M) \int {dz \over z^2} D_b^h(z,M_F) \nn \\
&&\int
{d\varphi_4 \over 2 \pi } \ {1 \over (4 \pi)^2}\  {1 \over 2 xS}\
{\ell^{\mu\nu} t_{\mu\nu}^{ab} \over q^4} \ dps' \eea

\noi where the phase space $dps'$ no longer contains parton 4
which fragments into $h(P_4)$. ($\eta_4$ is the pseudo-rapidity of 
the observed hadron).\par

It is useful to give a more explicit form to the tensor product
in the $\gamma^*-p$ frame
by defining the transverse polarization
vectors $\varepsilon_1^{\mu} = (0,1,0,0)$, $\varepsilon_2^{\mu} =
(0,0,1,0)$ and the scalar polarization vector $\varepsilon_s^{\mu} = {1
\over \sqrt{Q^2}} (q^z, 0, 0, q^0)$ with $q^{\mu} = (q^0, 0, 0, q^z)$
the virtual photon momentum

\bea
\label{5e}
&&\ell^{\mu\nu} t_{\mu\nu} = Q^2(t_{11} + t_{22}) + 4 \left ( {Q^2(1
- y) \over y^2} - m_e^2 \right ) t_{11}\nn \\
&&+ 4 {2 - y \over y} \ell_x \sqrt{Q^2} \ t_{s1} + Q^2 {4 (1 - y)
\over y^2} t_{ss} \ ,
\eea

\noi the transverse momentum $\ell_x$ of the initial lepton being 
along the $x$-axis. \par

In the limit $Q^2 \to 0$ and after azimuthal averaging over $\varphi_4$
we recover the
unintegrated Weizs\"acker-Williams expression

\beq \label{6e} {1 \over 2} \ {\ell^{\mu\nu} t_{\mu\nu} \over Q^4} =
\left ( {1 + (1 - y)^2 \over y Q^2} - {2y \ m_e^2 \over Q^4}\right )
\sigma_{\bot} +
{\cal O}\left ( \left ( Q^2\right )^0 \right ) \eeq

\noi with $\sigma_{\bot} = {1 \over 2y} \left ( t_{11} + t_{22}\right
)$. \par

Actually the limit (\ref{6e}) is correct only if $\lim\limits_{Q^2
\to 0} t_{ss} =
{\cal O}(Q^2)$. This is not true if an initial collinearity is present in
the partonic tensor (light partons are massless) which leads to the
behavior $\lim\limits_{Q^2 \to 0} t_{ss} = {\cal O}(1)$. This point is
discussed at the end of this section.\par

The partonic tensor is given by a perturbative expression in
$\alpha_s$. The Born contribution is of order ${\cal O}(\alpha_s)$ and
corresponds to the QCD Compton subprocess $\gamma^*+q \to g+q$ and the
fusion process $\gamma^* + g \to q + \bar{q}$. Higher Order ${\cal
O}(\alpha_s^2)$ corrections to the Born cross section have been
calculated in ref. \cite{12r}. In the course of these HO calculations a
resolved component appears, corresponding to subprocesses in which the
virtual photon creates a collinear $q$-$\bar{q}$ pair~; the quark or the
antiquark subsequently interacts with a parton of the proton.\par

Let us study this contribution in detail by considering the simple
model illustrated by the gauge invariant set of Feynman graphs
displayed in Fig. 1. The neutral parton of momentum $p$ is off-shell and is
part of a hard process also involving a parton of the proton. The final
parton of momentum $p_4$ fragments into the observed large-$E_{\bot}$
hadron of transverse energy $E_{\bot 4}$. All the results described
below can easily be obtained from the expressions given in appendix 1.

\begin{figure}[htb]
\vspace{9pt}
\centering
\includegraphics[width=5in,height=2in]{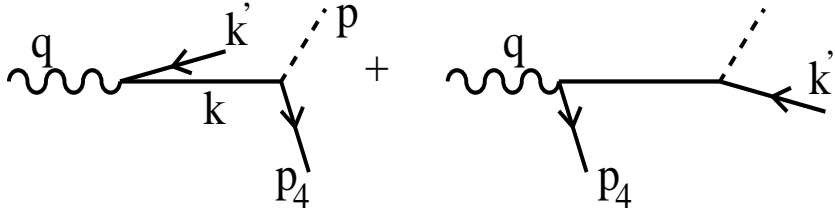}
\caption{Feynman graphs leading to a resolved contribution.}
\label{fig:1}
\end{figure}

The cross section corresponding to the graphs of Fig.~1 has double and
single poles in $k^2 = (q - k')^2$. The interference term between graphs
(a) and (b) has a single pole which leads to an expression 
proportional to $\ell n {p_{\bot 4}^2 \over
-q^2}$, after integration over
$k'_{\bot}$. However, a prefactor $q^2$ is present in all tensor components
 $t_{\scriptscriptstyle AB}$ ({\eightrm A, B}$ = S,1,2)$. As a 
result, these components have no
singularities when $q^2$ tends to zero. This well-known behaviour is due
to current conservation (for the components involving a scalar photon)
and to the fact that interference terms are not singular for transverse
photons. Therefore, let us concentrate on the square of graph (a) and
start with the transverse component which has the expression (after
integration over the azimuthal angle $\varphi_{k'})$

\beq
\label{7e}
t_{ii} = {\alpha \over 2\pi} \ 3 e_f^2\int dz \left [(1-z)^2 + z^2 
\right ] \int_{-q^2z}^{p_{\bot 4}^2 /(1-z)} {dk^2 \over k^4} \left \{ 
zq^2 - k^2  \right \} {|\mu (0)|^2 \over 2z} \quad (i = 1,2)
\eeq

\noi where $\mu(0)$ is the hard subprocess amplitude, here
representing the process $k + p \to p_4$, in which we have set $k^2$
and $k_{\bot}^2$ equal to zero. The upper limit of the
$k^2$-integration indicates the scale at which the collinear
approximation used in (\ref{7e}) by setting $k^2$ and $k_{\bot}^2$
equal to zero is no longer valid, using formulae (\ref{4e}) and 
(\ref{5e}), and after integration over $k^2$. The contraction with 
the leptonic
tensor leads to

\bea \label{8e} &&{2\pi d\sigma \over d\varphi dQ^2 dy} = {\alpha \over 2 \pi}
\left [ {1+(1-y)^2 \over y} \ {1 \over Q^2} - {2m_e^2y \over 
Q^4}\right ]  \int dz {\alpha \over 2\pi} \ 3e_f^2 [(1-z)^2 + z^2] 
\nn \\
&&\cdot \left \{ \ell n {p_{\bot 4}^2 \over Q^2} - \ell n\ z - \ell n 
(1-z) - 1
\right \} \widehat{\sigma}  \ ,\eea

\noi where we define $\widehat{\sigma} = {|\mu (0)|^2 \over 2 zyS}$ 
(we have not written the contribution of order ${\cal O}(Q^2/p_{\bot 
4}^2)$). This expression is the lowest order resolved cross section 
and is exactly the one which is obtained in the course
of the calculation of HO corrections to the direct Born terms \cite{12r}. The
Weisz\"acker-Williams distributions of the virtual photon in the initial
electron and the quark distribution in the virtual photon are universal
as they do not depend on the particular hard process described by
cross-section $\widehat{\sigma}$. Expression
(\ref{8e}) is the starting point of this paper. Indeed, when $\ell n
{p_{\bot 4}^2 \over Q^2}$ is large, one cannot content oneself with this
approximation and corrections of the type $\alpha_s^k \left ( \ell n
{p_{\bot}^4 \over Q^2}\right )^n$ with $k = n , n+1$ must be calculated
and resummed. These corrections modify expression (\ref{8e}) at the
Leading Order ($k=n$) and at the Next-to-Leading Order ($k =
n+1$) approximation.\par

In order to avoid double counting, expression (\ref{8e}) must be
subtracted from the NLO direct cross section. Actually the exact
expression to be subtracted is a matter of factorization scheme. We 
define the resolved component by

\beq
\label{9e}
{2\pi d \sigma^{res} \over d \varphi dQ^2dy} = {\alpha \over 2 
\pi}\left [ {1 + (1-y)^2 \over y} \ {1 \over Q^2} - {2m_e^2y \over 
Q^4} \right ] \int dz {\alpha \over 2 \pi} 3e_f^2 \left [ (1-z)^2 
+z^2\right ] \ell n {M_{\gamma}^2 \over Q^2} \widehat{\sigma}
\eeq

\noi where we introduce the factorization scale $M_{\gamma}$ with
$M_{\gamma} = {\cal O}(E_{\bot 4})$. After subtraction, the part of
(\ref{8e}) left in the direct HO corrections is obtained from
(\ref{8e}) by the substitution $\ell n {p_{\bot 4}^2 \over Q^2} \to \ell n
{p_{\bot 4}^2 \over M_{\gamma}^2}$. We call this factorization scheme
the {\it virtual factorisation scheme}. This is a natural scheme in
virtual photoproduction in which all the $\ell n\  Q^2$-terms are resummed
in the parton distributions. Then the total NLO cross section is given
by the sum of the subtracted direct cross section and of the resolved
cross section calculated at NLO at the scale $M_{\gamma}$. The
variations of the resolved cross section with $M_{\gamma}$ are partly
compensated by the $\ell n\  M_{\gamma}^2$ terms, these remain in the direct
cross section so that the total NLO cross section exhibits a smaller
sensitivity to $M_{\gamma}$ than the LO cross section.\par

Of course this procedure is useful as long as $p_{\bot 4}^2 \gg
Q^2$. Actually the collinear approximation used in (\ref{7e}) is valid
if $Q^2 \ \lsim \ k_{\bot}^2 \ \lsim \ p_{\bot 4}^2$, which allows us
to put $k_{\bot}^2 = 0$ in the hard cross section.  When
$p_{\bot 4}^2 \ \lsim\ Q^2$, this upper limit is incorrect. Let us
rewrite the $k^2$-integral in (\ref{7e}) in terms of $\vec{k}_{\bot}^2$

\beq
\label{10e}
\int_{-q^2z} {dk^2 \over |k^2|} \sigma (\vec{k}_{\bot}^2) = \int_0 {d
\vec{k}_{\bot}^2 \over k_{\bot}^2 - q^2z(1-z)} \ \sigma
(\vec{k}_{\bot}^2)\ .\eeq

\noi This integral is sensitive to the dependence on
$\vec{k}_{\bot}^2$ of the $2 \to 2$ subprocess cross section
$\widehat{\sigma} (\vec{k}_{\bot}^2)$ which behaves approximately 
like ${\cal O}({1
\over (k_{\bot} + p_{\bot 4})^2})$. This behavior shows that no
collinear logarithmic terms (coming from the denominator 
$\vec{k}_{\bot}^2 + Q^2z(1-z)$)
are present when $p_{\bot 4}^2 \ \lsim \ Q^2$. Therefore,
the resolved component must be proportional to the result of 
$k_{\bot}^2$-integration in which the upper limit $p_{\bot 4}^2$ is 
replaced by $Q^2 + p_{\bot 4}^2$.  For $p_{\bot 4}^2 \gg Q^2$ we have
the case already discussed and for $p_{\bot 4}^2 \ll Q^2$, there is no
resolved component. \par

As a consequence it is more appropriate to define
the factorization scale

\beq
\label{11e}
M_{\gamma}^2 = Q^2 + C_{\gamma}^2 E_{\bot 4}^2
\eeq

\noi ($E_{\bot 4}$ is the transverse energy of the observed hadron)
which has the following correct properties. 1) It does not depend on
kinematical variables internal to the subprocess which may lead to 
incorrect results
when HO corrections are calculated \cite{14r}~; 2) The resolved component
calculated at $M_{\gamma}$ vanishes when $Q^2 \gg
E_{\bot 4}^2$~; 3) again we find the conventional factorization scale
$M_{\gamma} \simeq C_{\gamma} E_{\bot 4}$ when $E_{\bot 4}^2 \gg Q^2$,
$C_{\gamma}$ being an arbitrary constant of order 1.\par

Let us finish this section by discussing the tensor components $t_{is}$
and $t_{ss}$ which come from the square of graph (a) in Fig. 1. The
components $t_{is}$ behave like $\sqrt{q^2} \ell n {p_{\bot 4}^2 \over -
q^2}$ and have no singularity at the limit $q^2 \to 0$. On the contrary
$t_{ss}$ has a constant behavior when $q^2 \to 0$

$$t_{ss} = {\alpha \over 2 \pi} 3e_f^2\int dz\ 4z(1-z) 
\int_{-q^2z}^{p_{\bot 4}^2/(1-z)} {d|k|^2 \over k^4} (-q^2) {|\mu 
(0)|^2\over 2} \ ,$$

\noi or

\beq
\label{12e}
t_{ss} = {\alpha \over 2 \pi} 3e_f^2 \int dz \ 4z(1-z) \left ( 1 - 
{-q^2z(1-z) \over p_{\bot 4}^2}\right ) {|\mu (0)|^2 \over 2z} \ ,
\eeq

\noi a result which leads to the scalar cross section

\beq
\label{13e}
{2\pi d\sigma^{scalar} \over d \varphi dQ^2dy} = {\alpha \over 2 
\pi}\  {2(1-y ) \over y}\ {1 \over Q^2} \int dz \ {\alpha \over 2 
\pi} \ 3e_f^2\ 4z(1-z) \widehat{\sigma}\ .
\eeq
\noi In going from (\ref{12e}) to (\ref{13e}), we dropped the
$-q^2z(1-z)/p_{\bot 4}^2$ term which
depends, through $p_{\bot 4}$, on the detailed kinematics of the
subprocess.\par

We observe that $t_{ss}$ has a ``constant'' behavior when $q^2 \to 0$
due to the double pole of the cross section. Actually the limit $q^2
\to 0$ corresponds to a non perturbative region for the
$k^2$-integration. If instead of $|k^2_{min}| = - q^2z$ we set
$|k_{min}^2| \sim \Lambda_{QCD}^2$, we would obtain a vanishing
cross section when $q^2 \to 0$. A similar result is obtained if we
consider massive quarks (with $|k^2_{min}-m^2| = -q^2z  + {m^2 \over 
1- z}$). Therefore, for a physical
process and a real photon, there is no $t_{ss}$ contribution, as can 
be expected. \par

However, let us notice that for small values of $Q^2  \sim
\Lambda_{QCD}^2$, the resolved cross sections, as defined in (\ref{8e})
and (\ref{13e}), strongly depend on the way the $k^2$-integral is
regularized, different lower bounds produce different $z$-dependence,
and thus different physical results even when $E_{\bot 4}^2/Q^2$ is
large. This paradox is however solved by the HO correction to parton
distributions in the photon discussed in the next section. There we
shall see that the NLO parton distributions contain a term that cancels
the unwanted $z$-dependent contribution, up to a vanishing term when
$E_{\bot 4}^2/Q^2$ tends to infinity. Actually this result is true for
all $z$-dependent terms of collinear origin (related to the lower limit
of the $k^2$-integration) present in (\ref{8e}) and (\ref{11e}). As a
consequence the scalar cross section (\ref{13e}) will be cancelled.

\section{NLO corrections} \hspace*{\parindent} In section 2 we defined
the resolved component of the transverse cross section $(i = 1,2)$.
(Here $\widehat{\sigma}^B$ is defined as the Born amplitude squared divided
by the flux factor $z$).

\beq \label{14e} t_{ii} = {\alpha \over 2 \pi} \ e_f^2\int dz \
P_{q\gamma }^{(0)}(z) \ell n {M_{\gamma}^2 \over  Q^2} \
\widehat{\sigma}^B \eeq

\noi where the factorization scale is given by (\ref{11e}) and
$P_{q\gamma}^{(0)}(z) = 3[z^2 + (1-z)^2]$ (for 1 quark species).
Expression (\ref{14e}) contains the lowest order (${\cal
O}(\alpha_s^0)$) parton distributions in the virtual photon

\beq \label{15e} q(z, M_{\gamma}^2,Q^2) = {\alpha \over 2 \pi}\ e_f^2\
P_{q\gamma}^{(0)} (z) \ell n {M_{\gamma}^2 \over Q^2}\ . \eeq

\noi The Born cross section $\widehat{\sigma}^B$ describes the
scattering between a quark of the virtual photon and a parton of the
proton producing two large-$p_{\bot}$ partons in the final state. \par

Leading Logarithm (LL) corrections, corresponding to the emission of
collinear gluons by the initial quark, can be obtained by solving the
following inhomogeneous DGLAP equation \cite{13r,14r} (we only reproduce the
evolution equation for the Non Singlet (NS) quark distribution $q_f^{NS}
= q_f + \overline{q}_f - \sum\limits_{f=1}^{N_f} (q_f +
\overline{q}_f)/N_f$) with $<e_f^2> = \sum\limits_f e_f^2/N_f$

\beq \label{16e}
M^2 \ {\partial q_f^{NS}(M^2,z) \over \partial M^2} =
{\alpha \over 2 \pi} \ 2[e_f^2 - <e_f^2>]\ P_{q\gamma}^{(0)} (z) + 
{\alpha_s(M^2) \over 2 \pi} \int_z^1 {dz'\over z'}\
P_{qq}^{(0)}\left ( {z \over z'} \right )  q_f^{NS}(M^2,z')  \eeq

\noi where $P_{qq}^{(0)}(z) = C_F ((1 + z^2)/(1-z))_+$. The lowest
order expression (\ref{15e}) is solution of such an equation when
$P_{qq}^{(0)}$ is set equal to zero. The solution of (\ref{16e}) for
the moments $q_f^{NS}(M^2, Q^2,n) = \int_0^1 dz\ z^{n-1} \
q^{NS}(M^2,Q^2,z)$ is given by ($d(n) = 2 P_{qq}^{(0)}(n)/\beta_0$ 
and $\beta_0$ is the lowest order coefficient of the $\beta$-function 
expansion ${\partial \alpha_s \over \partial \ell n (\mu^2)} = \beta 
(\alpha_s) \cong - {\alpha_s^2 \over 4 \pi} \beta_0$)

\beq \label{17e} q_f^{NS,LL}(M^2,Q^2,n) = {4 \pi \over 
\alpha_s(M^2)}\ {\alpha \over 2 \pi}\ {2(e_f^2 - <e_f^2>) 
P_{q\gamma}^{(0)}(n) \over \beta_0(1 - d(n))} \left ( 1 - \left ( 
{\alpha_s(M^2) \over \alpha_s(Q^2)}\right )^{1-d(n)}\right ) \eeq

\noi with the boundary condition $q_f^{NS,LL}(Q^2, Q^2, n) = 0$. The Leading
Logarithm NS expression for the resolved cross section is now given by
\beq \label{18e}
t_{ii}^{LL} = q_f^{NS,LL} \otimes \widehat{\sigma}^B \ ,
\eeq

\noi which is expression (\ref{14e}) in which the lowest order parton
distribution is replaced by the LL solution (\ref{17e}). \par

The next step is to look for a Next to Leading Order (NLO)
expression for $t_{ii}$, which requires the calculations of HO
corrections to both $q^{LL}$ and $\widehat{\sigma}^B$. Indeed the
structure of these HO corrections is the following. The hard cross
section has the expression ($\widehat{\sigma}^B$ is of order ${\cal
O}(\alpha_s^2)$)

\beq \label{19e} \widehat{\sigma}^{NLO} = \widehat{\sigma}^B +
\alpha_s^3 B \eeq

\noi whereas the parton distributions behave, in the asymptotic domain
$M_{\gamma}^2/Q^2 \gg 1$, like

\beq \label{20e} q^{NLO} = {a \over \alpha_s(M_{\gamma}^2)} + b\ . \eeq

\noi It is clear from (\ref{19e}) and (\ref{20e}) that a NLO expression
for $t_{ii}$ can only be obtained by calculating both
$\alpha_s^2B$ and $b$.

\subsection{The hard resolved cross section at NLO}
\hspace*{\parindent} The calculation of the HO corrections to the 
hard resolved cross
section is the simpler part of the NLO program, since these HO are
the same in real and virtual photoproduction reactions, providing we 
work in the same factorization scheme.
Therefore, we can borrow the results of ref. \cite{15r} obtained for the
real photoproduction of large $E_{\bot}$ hadrons. \par

Let us elaborate this point by first studying the resolved Born term. 
In the real case, instead of
(\ref{7e}) we obtain  the following expression

\beq \label{21e} t_{ii} = {\alpha \over 2\pi} \ 3 e_f^2\int dz \left
[ z^2 + (1-z)^2 - \varepsilon  \right ] {(4 \pi \mu^2)^{\varepsilon} \over (1
- \varepsilon )} \ {\Gamma (1 - \varepsilon) \over \Gamma (1 -
2\varepsilon)} \int_{0}^{p_{\bot 4}^2} {dk_{\bot}^2 \over
(k_{\bot}^2)^{1 + \varepsilon}} \widehat{\sigma}_{\varepsilon}^B  \eeq

\noi in which we use the dimensional regularization and $n = 4 - 2
\varepsilon$. The expression between the square brackets is the
$n$-dimensional DGLAP branching function~; the factor
$(4\pi)^{\varepsilon}\Gamma (1 - \varepsilon)/(\Gamma(1 - 2\varepsilon
) (1 - \varepsilon ))$ comes from the azimuthal integration and the
$n$-dimensional photon spin average. After integration over
$k_{\bot}^2$, we obtain $(1/\overline{\varepsilon} = {1 \over
\varepsilon} + \ell n \ 4 \pi - \gamma_E)$

\bea \label{22e} &&t_{ii} = {\alpha \over 2\pi} \ e_f^2\int dz \left (  - {1
\over   \overline{\varepsilon} } + \ell n {M_{\gamma}^2 \over \mu^2}  \right )
P_{q\gamma}^{(0)}(z) \widehat{\sigma}_{\varepsilon}^B + {\alpha \over 
2\pi} \ e_f^2\int dz\nn \\
&&\left ( \ell n {p_{\bot 4}^2 \over M_{\gamma}^2} P_{q\gamma}^0(z)
 - P_{q\gamma}^0(z) + 3\right ) \widehat{\sigma}^B \eea

\noi where the limit, when $\varepsilon$ tends to zero, of the $n$-dimension
Born cross section $\widehat{\sigma}_{\varepsilon}^B$ is simply 
$\widehat{\sigma}^B$ of
expression (\ref{14e}). This expression is identical to that obtained 
in the calculation of the HO corrections to the real direct term.\par

At this point, if we subtract the term proportional to $\left ( - {1 \over
\varepsilon} + \ell n {M_{\gamma}^2 \over \mu}\right )$ from (\ref{22e}), which
defines the $\overline{\rm MS}$ factorization scheme, we obtain a
direct HO subtracted contribution different from the one found in the
virtual case (cf. expressions (\ref{8e}) and (\ref{9e})). However, as
we shall see in the next subsection, this scheme dependence is
compensated by the NLO corrections to the parton distributions.\par

Now let us go one step further and consider ${\cal O}(\alpha_s)$ 
corrections to the resolved expression (\ref{9e}). These HO 
corrections are the same in the real and in the virtual case, with 
the exception of collinear contributions coming from
the branching $\gamma^* \to q + \overline{q} + g$ and containing 
$(\ell n\  p_{\bot 4}^2/Q^2)^n$ $(n = 1,2)$ terms. These logarithmic
terms can be factorized and resummed at the NLO approximation with the
result (we consider only the Non Singlet case)

\beq \label{23new}
q_{\gamma}^{NLO}(M_{\gamma}^2,Q^2) \otimes \left ( 1 + {\alpha_s 
\over 2 \pi} h_{qq} + {\alpha_s \over 2 \pi} \ P_{qq}^{(0)} \ell n 
{p_{\bot 4}^2 \over M_{\gamma}^2}\right ) \otimes \widehat{\sigma}^B
\eeq

\noi where $\otimes$ indicates convolutions in the longitudinal
variable. The factor 1 in the parenthesis corrsponds to the Born 
contribution (\ref{9e}). The term ${\alpha_s \over 2 \pi} h_{qq} 
\otimes
\widehat{\sigma}^B$ is the collinear HO correction calculated in the 
virtual factorization scheme
(resummation of all the $\ell n\  Q^2$ terms in the parton distribution
function with the boundary condition $q_{\gamma}^{NLO} (Q^2,Q^2) = 0$).
However the physical (direct + resolved) cross section is factorization scheme
invariant and can be written in terms of the $\overline{\rm MS}$ quantities
$\overline{q}_{\gamma}^{NLO}$ and $\overline{h}_{qq}(z)$. As a result we can
use the HO correction calculated in ref. \cite{15r} in the $\overline{\rm MS}$
scheme if we also use parton distributions (and a direct term) 
calculated in the same
scheme.\par

The authors of ref. \cite{11r,13new,14new} also worked in the $\overline{\rm
MS}$ scheme in their study of the electroproduction of large-$p_{\bot}$
jets, and they established the expression which must be subtracted from
the virtual direct term in order to obtain the $\overline{\rm MS}$
direct term. We comment on their results at the end of section 3.2.

\subsection{The virtual parton distributions at NLO}
\hspace*{\parindent} In order to delimit the problem of the 
Factorization Scheme (FS) in the
virtual parton distributions, we study the simple case of the DIS on a
virtual photon and we consider the $n$-moment of the structure 
function ${\cal F}_2^{\gamma}= F_2^{\gamma}(x, K^2,
Q^2)/x$ in which $Q^2 = |q^2|$ is the virtuality of the target photon,
$K^2 = |k^2|$ is the virtuality of the probe photon and $x$ the 
Bjorken variable. To make the
connection with the transverse cross section defined in (\ref{7e}),
${\cal F}_2^{\gamma}$ is defined by an average over the transverse spin of the
target photon only. To simplify the discussion we only consider the Non
Singlet contribution. ${\cal F}_2^{\gamma}$ is the sum
of a resolved part and a direct part (we drop the indices $n$)

\beq \label{23e} {\cal F}_2^{\gamma}(K^2, Q^2) = C_{2,q}(\alpha_s (K^2))
\cdot q_{\gamma}^{NS}(K^2, Q) + C_{2,\gamma}^{NS}(\alpha_s(K^2)) \ . \eeq

\noi In (\ref{23e}) ${\cal F}_2^{\gamma}$ is proportional to
$\sum\limits_{f=1}^{N_f} e_f^2 [e_f^2 - <e^2>]$~; we drop this factor
which is useless in the present discussion. The direct part,
$C_{2,\gamma}^{NS}$, and the resolved hard cross section $C_{2,q}$ (the
Wilson coefficient) are expansions in $\alpha_s(K^2)$. All the $\ell n \
Q^2$ dependent terms are collected in the virtual quark distribution
$q_{\gamma}^{NS}$ with the boundary condition 
$q_{\gamma}^{NS}(Q^2,Q^2) = 0$. This defines the virtual 
factorization scheme already
mentioned in section 2. In fact (\ref{23e}) is the final result 
obtained by Uematsu and Walsh \cite{10r} in their study of the 
virtual photon structure function, using the OPE and the MS 
factorization scheme as a starting point. The distribution 
$q_{\gamma}^{NS}$ verifies the
inhomogeneous DGLAP equation (\ref{16e}) in which the lowest order
branching function $P_{q\gamma}^{(0)}$ and $P_{qq}^{(0)}$ must be
replaced by the all order functions
$P_{q\gamma}^{NS}(\alpha_s(M_{\gamma}^2))$ and $P^{NS} (\alpha_s
(M^2))= {\alpha_s (M^2) \over 2 \pi} P_{qq}^{(0)} + \cdots$ which are 
expansions in $\alpha_s(M^2)$ and depend on the
factorization scheme. The solution of (\ref{16e}) can be written 
(from now on we drop the index NS)

\beq \label{24e} q_{\gamma} (K^2, Q^2) = {\alpha \over 2 \pi}
\int_{\alpha_s(Q^2)}^{\alpha_s(K^2)} {d \alpha '\
P_{q\gamma}(\alpha ') \over \beta (\alpha
')} \ e^{\int_{\alpha '}^{\alpha_s(K^2)}
{d\alpha '') \over \beta (\alpha
'')} P (\alpha '')} \ . \eeq

${\cal F}_2^{\gamma}$, being a physical observable, must be FS scheme invariant
and cancellation must exist in (\ref{23e}) between the various scheme dependent
contributions. Let us first note that $C_{2,\gamma}(K^2)$ is FS 
scheme invariant because

\beq
\label{25new}
{\cal F}_{\gamma}^2(Q^2,Q^2) = C_{2,\gamma}(\alpha_s(Q^2)) \ .
\eeq

To study the scheme dependence of $q_{\gamma}$, let us start from
expression (\ref{24e}) and define a new DGLAP branching function
$\overline{P}$ by

\beq \label{25e} P = \overline{P} - \delta P \eeq

\noi where $\delta P$ is an arbitrary expansion in $\alpha_s$
starting at order ${\cal O}(\alpha_s^2)$

\beq \label{26e} q_{\gamma}(K^2, Q^2) = {\alpha \over 2 \pi}\ e^{-
\int_0^{\alpha_s(K^2)} {d \alpha '' \over \beta (\alpha '')} \delta
P(\alpha '')}\cdot \widetilde{q}_{\gamma}(K^2,Q^2)  \eeq

\noi with $\widetilde{q}_{\gamma}$ given by

\beq \label{27e} \widetilde{q}_{\gamma}(K^2, Q^2) \equiv
\int_{\alpha_s(Q^2)}^{\alpha_s(K^2)} {d \alpha '
\over \beta (\alpha ')} \left [ P_{q\gamma}(\alpha ') e^{\int_0^{\alpha
'} {d \alpha '' \over \beta (\alpha '')} \delta P(\alpha
'')}\right ] \cdot e^{\int_{\alpha '}^{\alpha_s(K^2)} {d \alpha ''
\over \beta (\alpha '')} \overline{P} (\alpha '')} \ .\eeq

\noi We see that the variation $\delta P$ can be absorbed in
$C_{2,q}$ (the hard resolved subprocess) and $P_{q\gamma}$, thus 
defining new expansions in
$\alpha_s$, $\overline{C}_{2,q} (\alpha_s(K^2))$ and
$\widetilde{P}_{q\gamma}(\alpha ')$, whereas ${\cal F}_2^{\gamma}$ is kept
unchanged

\beq \label{28e} {\cal F}_2^{\gamma}(K^2,Q^2) =
\overline{C}_{2,q}(\alpha_s(K^2)) \widetilde{q}_{\gamma}(K^2, Q^2)
+ C_{2, \gamma} (\alpha_s(K^2))\ . \eeq

\noi Let us now study the effects of modifying $\widetilde{P}_{q\gamma}$

\beq \label{29e} \widetilde{P}_{q\gamma} = \overline{P}_{q\gamma} -
\delta P_{q\gamma} \eeq

\noi with the arbitrary series $\delta P_{q\gamma}$ starting at order 
${\cal O}(\alpha_s)$

\beq \label{30e} \widetilde{q}_{\gamma}(K^2, Q^2) = 
\overline{q}_{\gamma}(K^2,Q^2)  - {\alpha \over 2\pi} 
\int_{\alpha_s(Q^2)}^{\alpha_s(K^2)}
{d \alpha ' \over \beta}  \delta P_{q\gamma}\
e^{\int_{\alpha '}^{\alpha_s(K^2)} {d \alpha '' \over \beta}
\overline{P}} \eeq

\noi where the parton distribution
$\overline{q}_{\gamma}$ is calculated in the bar-scheme

\beq \label{31e} \overline{q}_{\gamma}(K^2, Q^2) = {\alpha \over 2 \pi}
\int_{\alpha_s(Q^2)}^{\alpha_s(K^2)} {d\alpha ' \over \beta}
\overline{P}_{q\gamma} \ e^{\int_{\alpha '}^{\alpha_s(K^2)} {d \alpha ''
\over \beta} \overline{P}}\ . \eeq

Finally for ${\cal F}_{\gamma}^2$ we obtain the expression

\beq \label{32e} {\cal F}_2^{\gamma}(K^2,Q^2) =
\overline{C}_{2,q}(\alpha_s(K^2)) \left ( \overline{q}_{\gamma}(K^2, 
Q^2) + q_{\gamma}^B(K^2, Q^2)\right )
+ \overline{C}_{2, \gamma} (\alpha_s(K^2)) \eeq

\noi where

\beq \label{33e} \overline{C}_{2, \gamma}(\alpha_s(K^2)) =
C_{2,\gamma}(\alpha_s(K^2)) - \overline{C}_{2, q}(\alpha_s(K^2)) 
{\alpha \over 2 \pi}
\int_0^{\alpha_s(K^2)} {d \alpha ' \over \beta } \delta P_{q\gamma} \
e^{\int_{\alpha '}^{\alpha_s(K^2)} {d \alpha '' \over \beta}
\overline{P}} \eeq

\noi and

\beq \label{34e} q_{\gamma}^B(K^2,Q^2) =
 {\alpha \over 2 \pi} \int_0^{\alpha_s(Q^2)} {d \alpha ' \over \beta }
\delta P_{q\gamma}\ e^{\int_{\alpha '}^{\alpha_s(K^2)} {d \alpha ''
\over \beta} \overline{P}} \ . \eeq

\noi Therefore in the new factorization scheme (the bar-scheme), the
structure of the expression for ${\cal F}_2^{\gamma}$ is the same as in
the original scheme, but the parton distribution does not vanish at
$K^2 = Q^2$ since $q_{\gamma}^B(Q^2,Q^2)$ is different from zero.
Therefore, by going from the virtual FS to the bar-scheme, {\it we find the
boundary condition that the bar-distribution must verify}. By rewriting
(\ref{34e}) as

\beq \label{37new} q_{\gamma}^B(K^2,Q^2) = 
e^{\int_{\alpha_s(Q^2)}^{\alpha_s(K^2)} {d \alpha '
\over \beta (\alpha ')} \overline{P}}\ {\alpha \over 2 \pi}
 \int_0^{\alpha_s(Q^2)} {d \alpha ' \over \beta (\alpha ')}
\delta P_{q\gamma}\ e^{\int_{\alpha '}^{\alpha_s(Q^2)} {d \alpha ''
\over \beta (\alpha '')} \overline{P}} \ , \eeq

\noi we see that $q_{\gamma}^B(K^2,Q^2)$ verifies the homogeneous
DGLAP equation and that the boundary condition is given, at the lowest
order ($\overline{P} = {\alpha_s \over 2 \pi} P_{qq}^{(0)}$, $\beta 
(\alpha_s ) = -
{\alpha_s^2 \over 4 \pi} \beta_0$ and $\delta P_{q\gamma} = {\alpha_s 
\over 2 \pi}\delta P_{q\gamma}^{(1)}$), by

\beq \label{35e} q_{\gamma}^B(Q^2, Q^2) = -  {\alpha \over 2
\pi} \ {\delta P_{q\gamma}^{(1)} \over P_{qq}^{(0)}} \ . \eeq

The bar-scheme can be any scheme, but it is convenient to work in
the $\overline{\rm MS}$ factorization scheme in which the two-loop
branching functions $P_{q\gamma}^{NS}$ and $P^{NS}$ are known. Moreover, in
the electroproduction of large-$p_{\bot}$ hadrons we also know the NLO
resolved subprocess cross section (the equivalent of $\overline{C}_{2,q}$)
calculated in the $\overline{\rm MS}$ scheme in ref. \cite{15r}. It is easy
to obtain $\delta P_{q\gamma}^{(1)}$ from expression (\ref{33e})
written at the lowest order on $\alpha_s$

\beq
\label{39new}
\overline{C}_{2, \gamma} - C_{2 , \gamma} =  {\alpha \over 2 \pi} \ 
%{\delta_{\gamma p}^{{\textstyle P}(1)} \over P_{qq}^{(0)}} \ .
{\delta P_{q\gamma}^{(1)} \over P_{qq}^{(0)}} \ .
\eeq

\noi $\overline{C}_{2, \gamma}$ is the $\overline{\rm MS}$ direct term
\cite{17r} and $C_{2,\gamma}$ is the virtual-scheme
direct\footnote{This direct term corresponds to transversely polarized
photons whereas the expression of ref. \cite{10r} also contains the
scalar contribution.} term \cite{10r}

\beq
\label{40new}
C_{2 , \gamma} = {\alpha \over 2 \pi} 6\left \{ 8x(1-x) - 2 +  \left 
( x^2 + (1-x)^2\right ) \ell n {1 \over x^2}\right \} \ ,
\eeq

\noi which leads to

\beq
\label{41new}
q_{\gamma}^B(Q^2,Q^2) = - {\alpha \over 2 \pi} 6 \left \{ \left ( x^2 
+ (1-x)^2\right ) \ell n [x(1-x)]+ 1 \right \} \ .
\eeq

Let us finish this section by going back to the direct cross section 
of large-$p_{\bot}$ hadron electroproduction. The $\overline{\rm MS}$ 
boundary condition can be obtained by
comparing expression (\ref{8e}) calculated in the virtual case and
expression (\ref{22e}) corresponding to the real case. We see that
subtracting $(z^2 + (1 - z)^2) \ell n  {1 \over z(1-z)} - 1$ from the 
virtual expression, we
find the $\overline{\rm MS}$ expression $1 - (z^2 + (1 - z))$. The 
term that we subtracted
is equal to (\ref{41new}) as can be expected\footnote{The subtraction 
term established in ref. \cite{11r} is
identical to the one found here except for a term proportional to $[z^2 +
(1-z)^2]\ell n\ z$. Therefore it does not totally ensure the
transformation from the virtual scheme to the $\overline{\rm MS}$ scheme
(for instance from $C_{2, \gamma}$ to $\overline{C}_{2, \gamma}$ in the
DIS case).}.

\subsection{The scalar parton distribution at HO} \hspace*{\parindent}
In section 2 we found a scalar resolved contribution (\ref{13e}) to the
electroproduction cross section corresponding to the scalar distribution
$q_0^{S}(z) = {\alpha \over 2\pi} 3e_q^2 [4z(1-z)]$. HO corrections to this
distribution correspond to the Feynman graph\par

\begin{figure}[htb]
\vspace{9pt}
\centering
\includegraphics[width=3in,height=2in]{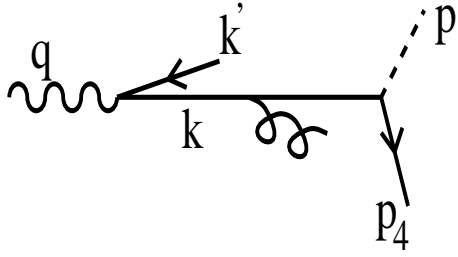}
\caption{A higher order correction to the Feynman graphs of fig. 1.}
\label{fig:2}
\end{figure}

\noi  of Fig.~2 (with an extra gluon in comparison to Fig.~1). 
Working in the LL approximation
and considering only terms proportional to $\ell n\  K^2/Q^2$), we have

\beq \label{36e} q_1^{S}(x,K^2,Q^2) = \int dz \ q_0^{S}(z) \int dz'\ 
{\alpha_s \over 2 \pi} \ P_{qq}^{(0)}(z')
\ \delta (zz' - x) \ell n\  K^2/Q^2 \ , \eeq

\noi or $q_1^{S}(n,K^2,Q^2) = q_0^{S}(n) {\alpha_s \over 2 \pi} 
P_{qq}^{(0)}(n) \ell n K^2/Q^2$. The
full LL expression $q^{S}(n) = \sum\limits_{k=1}^{\infty} q_k^{S}(n)$ is easily
resummed

\beq \label{37e} q^{S}(n,K^2,Q^2) = \int_{\alpha_s(Q^2)}^{\alpha_s(K^2)} {d
\alpha ' \over \beta ( \alpha ')} q_0^{S}(n) \ {\alpha ' \over 2 \pi} 
P_{qq}^{(0)} (n)\
e^{\int_{\alpha '}^{\alpha_s(K^2)} {d \alpha '' \over \beta ( \alpha
'')} {\alpha ''_s \over 2 \pi} P_{qq}^{(0)}}\ . \eeq

\noi This solution is similar to expression (\ref{24e}), but with an
inhomogeneous branching function starting at order ${\cal
O}(\alpha_s)$~; it can be written

\beq \label{38e} q^{S}(n,K^2,Q^2) = - q_0^S(n) \left [ 1 - \left ( 
{\alpha_s(K^2) \over
\alpha_s(Q^2)}\right )^{-2P_{qq}^{(0)}/\beta_0} \right ]\ . \eeq

\noi Adding this contribution to the lowest order one (\ref{13e}), we
see that the latter is cancelled and replaced by a contribution
which vanishes asymptotically because of the factor
$(\alpha_s(K^2)/\alpha_s(Q^2))^{-2P_{qq}^{(0)}/\beta_0}$. Therefore the
scalar contribution (\ref{13e}) which is target dependent (it depends
on the regularization of the $k^2$-integral as discussed in section 2)
is cancelled. This mechanism is actually quite general and also valid for the
transverse component. \par

Therefore, in the electroproduction case a full treatment of the scalar
cross section amounts to subtract expression (\ref{13e}) from the NLO
direct cross-section and to add the scalar resolved component

\beq \label{45e} {d\sigma^{scalar} \over d \varphi dQ^2dy} = {\alpha
\over 2 \pi}\ {2(1-y) \over y}\ {1 \over Q^2}
\int dz\ q_{\gamma}^S(z, M_{\gamma}^2, Q^2) \widehat{\sigma} \eeq

\noi where we define\footnote{This parton distribution in the
scalar virtual photon has been studied in ref. \cite{18r,19r}.}

\beq \label{46e} q_{\gamma}^S(n,M_{\gamma}^2, Q^2) = q_0^S(n) \left (
{\alpha_s(M_{\gamma}^2) \over \alpha_s (Q^2)}\right
)^{-2P_{qq}^0/\beta_0} \ . \eeq

\section{NLO parametrization of the virtual photon structure function}
\hspace*{\parindent}
Although the works of Uematsu and Walsh \cite{10r} and Rossi \cite{20r}
date back to the eighties, interest in the
virtual photon structure function grew much later, thanks to the HERA 
experiments.
Since then several papers have been published discussing the parton 
distributions
in virtual photons at the LL approximation \cite{21r,22r,23r} and NLO
approximation \cite{24r,25r,26r}, and emphasizing the possibility to
measure them in electroproduction experiments 
\cite{13new,14new,27r,28r,32new,35new,29r}.
In this section we present a study of the parton distributions in
virtual photons performed in the $\overline{\rm MS}$ scheme, using the
results of section 3 on the inputs at $K^2
= Q^2$. We choose $Q^2$ large enough to neglect non perturbative
effects, and therefore we do not present results on the limit $Q^2 \to
0$. A NLO study has also been done by authors of refs. \cite{24r} and 
\cite{25r} in
the DIS$_{\gamma}$ scheme with emphasis the real limit $Q^2
\to 0$. We may note however that the solutions of \cite{25r} do not 
fulfil condition (\ref{25new}) of
section 3.\par

The parton distributions are solutions of the same NLO inhomogeneous
differential equations as in the real case. The NS equation is given by
expressions (\ref{16e}) in which ${\alpha \over 2\pi}
P_{q\gamma}^{(0)}$ and ${\alpha_s \over 2 \pi} P_{qq}^{(0)}$ must be
replaced by NLO (two loops) DGLAP kernels~; the equations of the
singlet sector as well as the expressions
of the kernels can be found in ref. \cite{16r}. The only change with 
respect to the real case is the
starting point of the evolution, $Q^2$ instead of $Q_0^2 \sim
(.5)^2$~GeV$^2$ in the real case. The boundary condition for the
distributions are given by expression (\ref{41new}) for the quark
distributions and we have $g_{\gamma}^B(Q^2,Q^2) = 0$ for the gluon
distributions. Another difference from the real case is the existence
of a scalar contribution. The latter has been discussed in section 3.3.\par

In section 3 we studied the massless case $m_q^2 \ll Q^2$. However
$\sqrt{Q^2}$ can be smaller than the bottom mass $m_b$ and we have to
determine what are the relevant boundary conditions for the bottom
quark distribution in the virtual photon. To find these, let us
introduce in expression (\ref{7e}) the kinematic corresponding to the
case in which the photon interacts with a massive quark. \par

\bea
\label{47e}
&&3 \int_{Q^2z + {m^2 \over 1-z}}^{p_{\bot 4}^2/(1-z)} {d|k^2 - m^2| 
\over (k^2 - m^2)^2} \left \{ |k^2 - m^2| {z^2 + (1-z)^2 \over z} + 
{2m^2z - Q^2z(z^2 + (1 - z)^2) \over z}\right \} \nn \\
&&= {1 \over z} \left \{ P_{q\gamma}(z) \ell n {p_{\bot 4}^2 \over 
m^2 + Q^2z(1-z)} + 6z (1-z) - {3 Q^2 z (1-z) \over m^2 + Q^2z(1-z)} 
\right \} \ .
\eea

\noi We notice that for $m^2 = 0$, we find the massless corrections
already given in (\ref{8e}) which are associated with the virtual
factorization scheme.\par

For $Q^2 = 0$, we find

\beq
\label{48e}
{1 \over z} \left \{ P_{q\gamma}(z)\  \ell n {p_{\bot 4}^2 \over m^2} 
+ 6z(1-z)\right \}
\eeq

\noi which is the $\overline{\rm MS}$ correction (once the term $\ell n
{M_{\gamma}^2 \over m^2}$ is subtracted). Therefore, in the massive
case ($Q^2 = 0$) we directly work in the $\overline{\rm MS}$-scheme and
we do not have to modify the factorization scheme as discussed in
section 3. Therefore, there exists a transition between case $Q^2 \gg
m^2$ and case $Q^2 \ll m^2$ that we should study in detail.\par

Let us start from expression (\ref{47e}). When $Q^2 < m^2$, we
factorize $P_{q\gamma} \ell n {M_{\gamma^2} \over m^2}$ which is the
contribution given by the evolution equation starting at the scale
$m^2$. The rest is given by (without the $1/z$ prefactor)

\beq
\label{49e}
P_{q\gamma} \ell n  {p_{\bot 4}^2 \over M_{\gamma}^2}  + 6z(1-z) + 
MC_<(m^2,Q^2;z)
\eeq

\noi with

$$MC_<(m^2,Q^2;z) = P_{q\gamma} \ \ell n {m^2 \over m^2 + Q^2z(1-z)} 
- {3 \cdot Q^2 z (1-z) \over m^2 + Q^2z(1-z)} \ ,$$

 \noi namely the usual masless $\overline{\rm MS}$ correction and 
corrections in $Q^2/m^2$. When $Q^2 > m^2$, we factorize $P_{q\gamma} 
\ell n {M_{\gamma}^2
\over Q^2}$ and we obtain

\beq
\label{50e}
P_{q\gamma} \ \ell n {p_{\bot 4}^2 \over M_{\gamma}^2} + P_{q\gamma} 
\ell n {1 \over z(1-z)} + 6z(1-z) - 3 + MC_>(m^2,Q^2;z)
\eeq

\noi with

$$MC_>(m^2,Q^2;z) = P_{q\gamma} \ell n {Q^2 z (1 - z) \over m^2 + 
Q^2z(1-z)} + {3 \cdot m^2 \over m^2 + z(1-z)Q^2} \ .$$

\noi We recognize the massless corrections in the virtual scheme and 
a massive $m^2/Q^2$ correction. \par

However the same massive corrections 
$MC_{\displaystyle{\mathrel{\mathop {>}_{\textstyle <}}}} (m^2Q^2;z)$ 
appear in the
calculation of the inhomogeneous kernel $k_q^{(1)}$ as outlined in
section 3.3. When we add the resolved and the direct contributions,
these massive corrections are cancelled and are replaced by
$MC_<(m^2,Q^2;n)(\alpha_s(M_{\gamma})^2/\alpha_s(m^2))^{-2P_{qq}^{(0)}/
\beta_0}$ or by
$MC_>(m^2,Q^2;n)(\alpha_s(M_{\gamma}^2)/$\break \noindent 
$\alpha_s(Q^2))^{-2P_{qq}^{(0)}/
\beta_0}$ when $Q^2 > m^2$. In the latter case, we still have to add
$-[P_{q\gamma}(z) \ell n z(1-z)+3]$ to $MC_>(m^2,Q^2;z)$ to move to the
$\overline{\rm MS}$ scheme with the result

\beq
\label{51e}
MC_>(m^2,Q^2;z)  = P_{q\gamma} \ell n {Q^2 \over m^2 + Q^2z(1-z)} - 
{3Q^2 z(1-z) \over m^2 + Q^2 z(1-z)}
\eeq

\noi which is equal to $MC_<(m^2,Q^2;z)$ when $m^2 = Q^2$.\par

Therefore, we can summarize our treatment of the massive quarks in the
following way. First we assume that the $m^2/E_{\bot 4}^2$ corrections
are properly taken into account in the direct term (or that they are
negligible when $E_{\bot 4}^2 \gg m^2)$. Second, we work in the massless
$\overline{\rm MS}$ scheme and we take into account mass corrections
through the input of the quark distributions. Thus we have the
following inputs for $m_c^2 \leq Q^2 \leq m_b^2$ (up to charge factors)

\begin{eqnarray}
\label{52e}
&\left . \begin{array}{l}  u(x,Q^2)\\ d(x, Q^2)\\ s(x,Q^2) 
\end{array}\right \} &\sim - P_{q\gamma} \ \ell n \ x(1-x) - 3\nn \\
&c(x, Q^2) &\sim P_{q\gamma} \ \ell n \ {Q^2 \over m_c^2 + Q^2 
z(1-z)}- {3Q^2z(1-z) \over m_c^2 + Q^2 z(1-z)} \nn \\
&b(x,m_b^2) &\sim P_{q\gamma} \ \ell n \ {m_b^2 \over m_b^2 + Q^2 
z(1-z)} - {3Q^2 z(1-z) \over m_b^2 + Q^2 z(1-z)} \ .
\end{eqnarray}

\noi Whereas for $Q^2 \geq m_b^2$, we have the input

\beq
\label{53e}
b(x,Q^2)  \sim P_{q\gamma} \ \ell n \ {Q^2 \over m_b^2 +Q^2 z(1-z)} - 
{3Q^2z(1-z) \over m_c^2 + Q^2  z(1-z)}\ .
\eeq

With these inputs we obtain the distributions shown in Fig.~3. We have
chosen $Q^2 = 8$~GeV$^2$ and $M_{\gamma}^2 = 25$ which correspond to
average values of $Q^2$ and $M_{\gamma}^2 = Q^2 + E_{\bot 4}^2$ of the
H1 experiment \cite{8r}. The distributions calculated in the
$\overline{\rm MS}$ scheme increase for $x$ going to one as we can see
from Fig.~3. This increase is compensated however by the behavior of
the direct term which contains terms in $\ell n (1-z)$ that become
negative at large $z$. We also remark the effect of the massive input
(\ref{52e}) for the charm quark distribution.\par

\begin{figure}[htb]
\vspace{9pt}
\centering
\includegraphics[width=4in,height=3in]{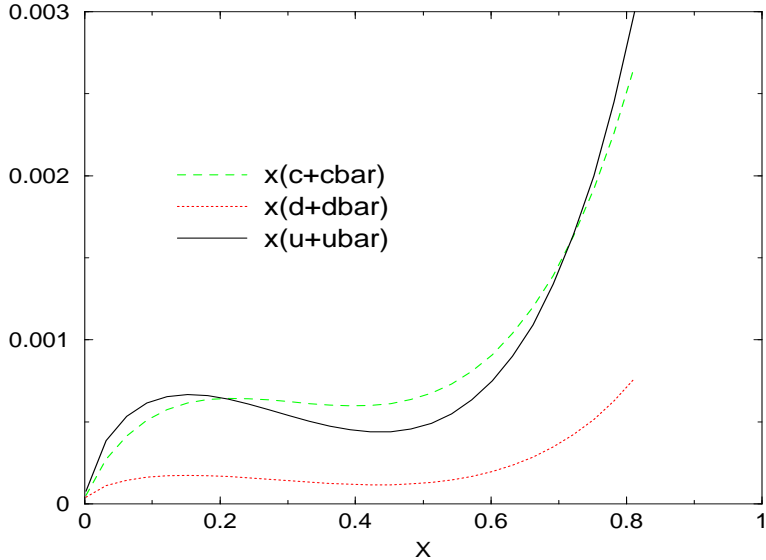}
\caption{The parton distributions in the virtual photon for $Q^2 = 
8$~GeV$^2$ and $M_{\gamma}^2 = 25$~GeV$^2$.}
\label{fig:3}
\end{figure}

We end this section by comparing our results with experimental data
obtained by the L3 collaboration \cite{4r} for the structure function

\beq \label{54e} F_{eff}^{\gamma} = F_{2 , \bot}^{\gamma} +
F_{2,s}^{\gamma} \eeq

\noi where the indices $\bot$ and $s$ refer to polarization of the 
target photon of
virtuality $Q^2$ (called $P^2$ in ref. \cite{4r}). In terms of these
components, the usual structure function $F_2^{\gamma}$ is written (the
tensor indices refer to the target current) $F_2^{\gamma} = - {1 
\over 2} g^{\alpha
\beta}(F_2^{\gamma})_{\alpha \beta} = F_{2, \bot}^{\gamma} - {1 \over
2} F_{2,s}^{\gamma}$. Until now we calculated only the transverse distributions
(Fig.~3)~; in order to obtain $F_{eff}^{\gamma}$ we have to add to
$F_{2, \bot}^{\gamma}$ the scalar contribution defined in expression
(\ref{46e}) for the quark component, in which a gluon distribution is 
also generated
by the DGLAP evolution equation. All our calculations are done in the
$\overline{\rm MS}$ scheme.\par

\begin{figure}[htb]
\vspace{9pt}
\centering
\includegraphics[width=4in,height=3in]{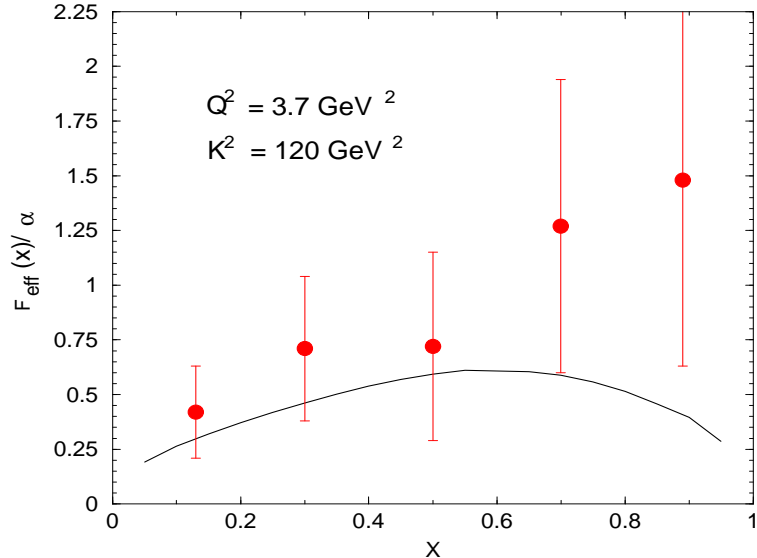}
\caption{The structure function $F_{eff}^{\gamma}/\alpha$ compared to 
L3 data \protect{\cite{4r}} with $Q^2 = 3.7$~GeV$^2$ and $K^2 = 
120$~GeV$^2$. Statistical and systematical errors are added linearly.}
\label{fig:4}
\end{figure}

In Fig.~4 we see that our predictions are in reasonable agreement with
data at low and medium values of $x$, but they undershoot them at large
values of $x$. Similar results have been obtained by the authors of
ref. \cite{12rnew}.

\section{Numerical results}
\hspace*{\parindent} We now turn to a phenomenological study of the 
Deep Inelastic
Production of large-$E_{\bot}$ hadrons. We concentrate mainly on the
resolved contribution studied in this paper and consider the H1 data \cite{8r}
already discussed in ref. \cite{12r} devoted to the direct
contribution. This allows us to make a connection between the results
presented here and those obtained in \cite{12r}. A more complete
phenomenological study of the new H1 data \cite{10rnew} will be presented
in a future paper \cite{32r}, in which we shall also discuss in 
detail the link between the present NLO cross section and the cross 
section based on the exchange of Reggeized gluons \cite{33new} in the 
$t$-channel \cite{15new,35r} \par

As in paper \cite{12r}, we use the MRST 99 (upper gluon) distributions
for the parton in the proton \cite{33r} and the KKP fragmentation
functions \cite{34r}. The strong coupling constant is given by an exact
solution of the two loop-renormalization group equation and we use
$\Lambda_{\overline{MS}}^{(4)}$ = 300 MeV. We take $N_f = 4$. Our 
calculations are
performed at $\sqrt{S} =$ 300.3 GeV and the forward-$\pi^0$ cross
section is defined with the following cuts. In the laboratory system a
$\pi^0$ is observed in the forward direction with $5^{\circ} \leq
\theta_{\pi^0} \leq 25^{\circ}$~; the laboratory momentum of the pion
is constrained by $x_{\pi^0} = E_{\pi^0}/E_P \geq .01$, and an extra
cut is put on the $\pi^0$ transverse momentum in the $\gamma^*-p$
center of mass system~: $E_{\bot \pi^0}^* > 2.5$~GeV. The inelasticity
$y = Q^2/x_{{\scriptscriptstyle B}_j}S$ is restricted to the range 
$.1 < y< .6$. We consider
only the contribution coming from transversely polarized virtual
photons, we shall comment briefly on the scalar contribution below.\par

Our numerical results obtained for the distribution 
$d\sigma/dx_{{\scriptscriptstyle B}_j}$
measured by H1 \cite{8r} in the range 4.5~GeV$^2 \leq Q^2 \leq
15$~GeV$^2$ are shown in Fig.~5. In order to shorten the numerical calculation
we do not integrate over $Q^2$, but instead use the average value of
$Q^2$, $<Q^2> = 8$~GeV$^2$, over the above range. We use the scale $Q^2
+ E_{\bot 4}^2$ in the entire series of calculations and we work in 
the $\overline{\rm MS}$
factorization scheme.

\begin{figure}[htb]
\vspace{9pt}
\centering
\includegraphics[width=4in,height=4in]{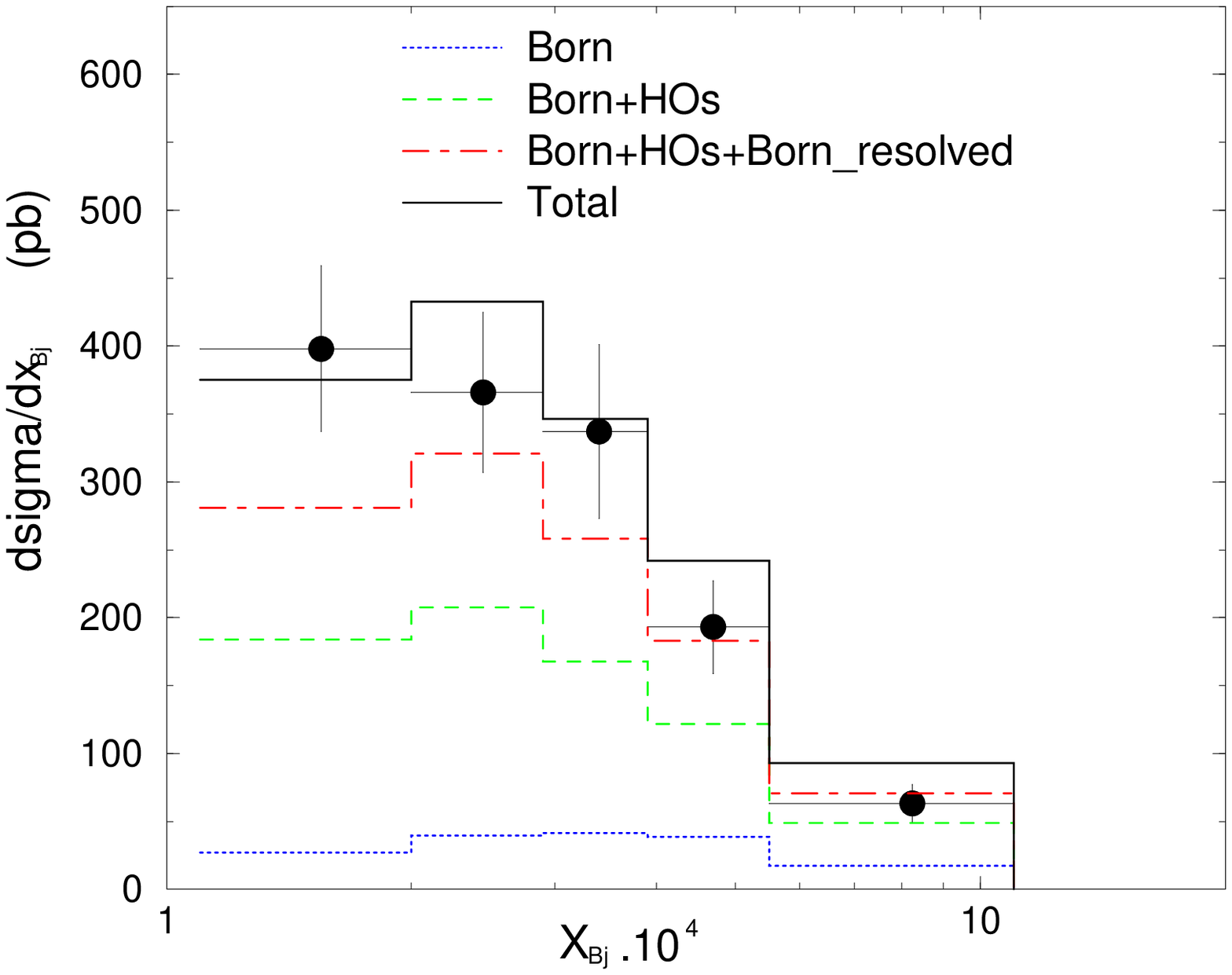}
\caption{The cross section $d\sigma/dx_{{\scriptscriptstyle B}_j}$ 
corresponding to the range $4.5$~GeV$^2 \leq Q^2 \leq 15$~GeV$^2$ 
compared to H1 data \protect{\cite{8r}}.}
\label{fig:5}
\end{figure}

 The direct HO corrections from which the resolved contribution is
subtracted, called HO$_s$, are different from those obtained in ref.
\cite{12r} in which we work in the virtual factorization scheme. In
both schemes they are very large. In ref. \cite{12r} we noticed that
the largest contribution to these corrections comes from the
subprocesses $\gamma^* + g \to g + q + \overline{q}$ and $\gamma^* + q
\to q + q + \overline{q}$. The sum of the HO$_s$ contributions and of
the resolved Born contribution should be factorization scheme
independent, up to ${\cal O} (\alpha_s)$ corrections. To check this
point, let us consider the bin $2.9 \cdot 10^{-4} \leq 
x_{{\scriptscriptstyle B}_j} \leq 3.9
\cdot 10^{-4}$. In ref. \cite{12r} we used the virtual factorization
scheme and we obtained $d\sigma/dx_{{\scriptscriptstyle B}_j} = 
155.5$~nb $+ 52.5$~nb =
208.0~nb for the sum. Note that the parton distributions used in that case are
simply the lowest order distributions (\ref{15e}) without QCD evolution.
In the $\overline{\rm MS}$ scheme we have 
$d\sigma/dx_{{\scriptscriptstyle B}_j} = 125.9$~nb
$+ 95.9$~nb = 221.8~nb, but we use the NLO parton distributions.
One can check that the small difference between the two sums
comes mainly from the gluon distributions not present in the lowest
order expression. If we only use the quark $\overline{\rm MS}$ NLO
distributions, we obtain $d\sigma^{Re}/dx_{{\scriptscriptstyle B}_j} 
= 83.1$~nb for the
resolved contribution and $d\sigma/dx_{{\scriptscriptstyle B}_j} = 
209$~nb for the sum. This
result shows that the QCD evolution is negligible in this kinematical
range (besides the generation of a small gluon distribution) and that
expression (\ref{15e}) gives a good description, in the virtual
factorization scheme, of the parton distributions in a virtual
photon. A similar observation has been made by the authors of ref. 
\cite{11r} in the case of jet production. Because of this small 
evolution, we also have $q_{\gamma}^2(n,
M_{\gamma}^2, Q^2) \simeq q_0^S(n)$. Therefore, it is not necessary to
subtract the scalar resolved component from the direct term and to
introduce a scalar (QCD evolved) resolved contribution. \par

The next point to observe from Fig.~5 is the importance of the HO
resolved corrections compared to the Born resolved
contributions, leading to a ratio $NLO/Born \simeq~2$ independent of
$x_{{\scriptscriptstyle B}_j}$. These large HO corrections correspond 
to a small value of
$E_{\bot 4}$ due to the small cut-off $E_{\bot \pi^0}^{*} \geq
2.5$~GeV. For a larger cut-off, for instance $E_{\bot \pi^0}^* \geq
5$~GeV, we obtain $NLO/Born = 1.65$ in the range $2.9 \cdot 10^{-4} \leq
x_{{\scriptscriptstyle B}_j} \leq 3.9\cdot 10^{-4}$.\par

The total cross section is in good agreement with data,
slightly overshooting them at $x_B \ \gsim\ 4\cdot 10^{-4}$, and little
room appears to be left for a BFKL-type contribution \cite{35r}. However
this last statement depends on the scale used in the calculation, here
$M = M_{\gamma} = M_F = \mu = (Q^2 + E_{\bot 4}^2)^{1\over 2}$, because
the cross section strongly depends on the scale $\mu$. In ref.
\cite{12r} we found that this was due to the importance of the
subprocesses $\gamma^* + g \to g + q + \overline{q}$ and $\gamma^* + q
\to q + q + \overline{q}$ with a gluon exchanged in the $t$-channel.
These processes correspond to the opening of new channels that are not present
at the Born level. They are of order ${\cal O}(\alpha_s^2)$ and
sensitive to the value of $\mu$ since there is no loop contributions at
this order to compensate the $\mu$-dependence. However, this remark is
not true for the resolved part of these subprocesses as soon as HO
corrections for the resolved cross section are calculated. For instance
the subprocess $\gamma^* + g \to g + q + \overline{q}$ contains a
resolved lowest order contribution $(\gamma^* \to q  \overline{q}) + g
\to g + q + \overline{q}$~; loop corrections to this contribution,
corresponding to HO corrections to the resolved cross section, generate
counter terms in $\ell n {E_{\bot 4}^2 \over \mu^2}$~; these in turn 
compensate the
$\mu$-dependence of the Born cross section. To check this point let us again
consider the bin $2.9\cdot 10^{-4} \leq x_{{\scriptscriptstyle B}_j} 
\leq 3.9\cdot 10^{-4}$. Keeping $M = M_{\gamma} = M_F = (Q^2 + 
E_{\bot 4}^2)^{1
\over 2}$ fixed and $E_{\bot\pi^0}^* > 5$~GeV, we vary $\mu =
C\sqrt{Q^2 + E_{\bot 4}^2}$ with $C$ ranging from .15 to 1.0. The
variations with $\mu^2$ of the Born and NLO resolved cross section are
shown in Fig.~6. \par

\begin{figure}[htb]
\vspace{9pt}
\centering
\includegraphics[width=4in,height=3in]{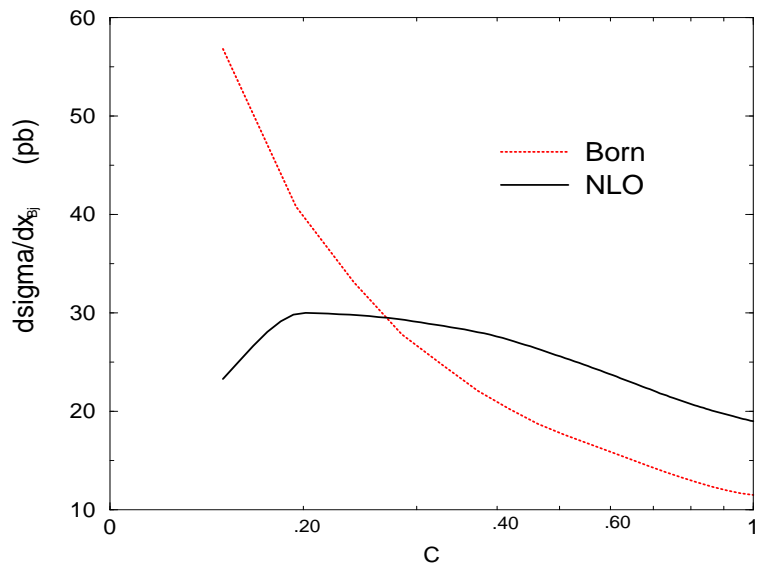}
\caption{The variation with $\mu = C \sqrt{Q^2 + E_{\bot 4}^2}$ of 
the resolved cross section.}
\label{fig:6}
\end{figure}

We see that the behavior of the Born cross section and that of the NLO
cross section are quite different. The latter has a maximum around $C
\simeq .2$ and is more stable with respect to the variations of $C$
than the Born contribution. This behavior does not occur for the NLO
direct contribution which always increases when $C$ decreases. Let us 
also note that this behavior cannot be observed for the cut
$E_{\bot \pi^0}^* > 2.5$~GeV. The HO corrections are too large and we
cannot reach a maximum of the NLO cross section, even for very small
values of $C$.\par

Let us conclude this section by noting another difference with respect
to the direct term in which the large contributions to the
forward cross section come from subprocesses involving the exchange of
one elementary gluon in the $t$-channel. In the resolved case, the
elementary gluon becomes reggeized, due to the HO corrections.
Therefore, the resolved cross section contains contributions
corresponding to the exchange of a reggeized gluon in the $t$-channel.

\section{Conclusion}
\hspace*{\parindent}
In this paper we calculated HO corrections to the resolved part of the
DIS cross section for the production of large-$E_{\bot} \pi^0$. This
involves the calculations of the HO corrections to parton distributions
in the virtual photon (of virtuality $Q^2$) and of the HO corrections
to the resolved subprocess.\par

We discuss the issue of the factorization scheme in detail and we
establish the inputs of the parton distributions in the $\overline{\rm
MS}$ scheme. Then the NLO parton distributions are obtained by solving
the DGLAP inhomogeneous evolution equation. They are confronted to LEP
data. \par

Our results for the NLO cross section are compared with H1 data for the
production of forward large-$E_{\bot \pi^0}$. We find a good agreement
with data, once the direct contribution is added to the resolved one.
This result is obtained for renormalization and factorization scales
equal to $Q^2 + E_{\bot \pi^0}^2$. In a study of the scale sensitivity,
we find that the resolved cross section is less sensitive to the
renormalization scale than the direct cross section. It is interesting
to notice that the authors of ref. \cite{14new} obtained very similar
results in their NLO study of the electroproduction of large-$E_{\bot}$
forward jets.\par

We conclude that the good agreement between the NLO calculations and
data leaves little room, in this kinematical range, for a BFKL-type contribution which resums a
ladder of reggeized gluon.

\section*{Acknowledgements}
\hspace*{\parindent}
I would like to thank P. Aurenche, R. Basu and R. Godbole for 
friendly and interesting discussions.

\newpage
\section*{Appendix 1}
\hspace*{\parindent} The square of the graphs in Fig. 1 possess single
and double poles in $k^2$. Actually, as explained in section 2, the
only relevant quantity is the square of graph (a) $S^{\mu\nu}/k^4$. Let
us isolate the hard subprocess amplitude $\mu$ by using the projector
defined in ref. \cite{37r} ${I\hskip -1 truemm P} = {({ / \hskip - 1.5 truemm
k} + m)][ {/ \hskip - 1.5 truemm \eta} \over 4k\cdot\eta}$ (with $\eta^{\mu}
= (1,0,0,1)$). On the left-hand side, this acts on the hard cross
section, and on the right-hand side, on the $\gamma^*q\bar{q}$
vertices~:

$${S^{\mu\nu}_{I\hskip - 1 truemm P} \over (eg)^2} = -
g^{\mu\nu} \left [ q^2 + (m^2 - k^2) {4 \eta \cdot q \over 4 \eta \cdot
k}\right ] |\mu |^2$$ $$+ 2 \left ( q^{\mu} k^{\nu} + q^{\nu} k^{\mu} -
2k^{\mu}k^{\nu}\right ) |\mu |^2$$ $$+ {m^2 - k^2 \over k \cdot \eta }
\left ( \eta^{\mu} (q^{\nu} - k^{\nu}) + \eta^{\nu}(q^{\mu} -
k^{\mu})\right ) |\mu |^2 \eqno({\rm A.1})$$

\noi with $|\mu|^2 = Tr \{\Gamma ({/\hskip - 2 truemm p} + {/ \hskip -
2 truemm k} + m) \Gamma ({/\hskip - 2 truemm k} + m)\}$ where $\Gamma$
describes the coupling of parton $p$ to the quark. All the expressions
discussed in section 2 can be obtained from (A.1) and the phase space
integration.

$$\int d^4k' \ \delta (k'^2 - m^2) = \int d^4k \ \delta \left ( (q-k)^2
- m^2\right ) = {1 \over 4} \int dk^2\ d \varphi_k dz$$

\noi with the definition $z = {k\cdot \eta \over q \cdot \eta} =
{k^{(-)} \over q^{(-)}}$ and the relation $\vec{k}_{\bot}^2 = -(k^2 -
m^2)(1-z) + q^2 z(1-z) - m^2$. The difference $(S^{\mu\nu} -
S_{{I\hskip - 1 truemm P}}^{\mu\nu})/k^4$ does not lead to singular
expressions when $q^2 \to 0$.

\end{document}